\documentclass[a4paper,10pt,twocolumn,aps]{revtex4}
\usepackage{graphicx}
\usepackage{dcolumn}
\usepackage{bm}
\usepackage{slashed}
\usepackage{amssymb}
\usepackage{amsmath}
\usepackage{amsthm}

\newtheorem{Theorem1}{Theorem}

\begin{document}

\title{Negativity for two blocks in the one dimensional Spin 1 AKLT model}

\author{Raul A. Santos }
\email{santos@insti.physics.sunysb.edu} 
\author{V. Korepin}
\email{korepin@max2.physics.sunysb.edu}
\affiliation{C.N. Yang Institute for Theoretical Physics,\\
       Stony Brook University,\\
  Stony Brook, NY 11794-3840, USA}
\author{Sougato Bose}
\email{sougato@theory.phys.ucl.ac.uk}
\affiliation{Department of Physics and Astronomy, University College London,
         Gower St., London WC1E 6BT, United Kingdom}

\begin{abstract}
In this paper we compute the entanglement, as quantified by negativity, between two blocks of length $L_A$ and $L_B$, separated by
$L$ sites in the one dimensional spin-1 AKLT model. We took the model with two different boundary conditions. We consider the case of
$N$ spins 1 in the bulk and one spin 1/2 at each boundary which constitute an unique ground state, and the case of just spins 1, even at the end
of the chain, where the degeneracy of the ground state is four.  In both scenarios we made a partition consisting of two blocks $A$ and $B$, 
containing $L_A$ and $L_B$ sites respectively. The separation of these two blocks is $L$. In both cases we explicitly obtain the reduced density 
matrix of the blocks $A$ and $B$. We prove that the negativity in the first case vanishes 
identically for $L\geq 1$ while in the second scenario it may approach a constant value $N=1/2$ for each degenerate eigenstate depending on the way one constructs
these eigenstates.  However, as there is some freedom in constructing these eigenstates, vanishing entanglement is also possible in the latter case. Additionally, we also 
compute the entanglement between non-complementary blocks in the case of periodic boundary conditions for the spin-1 AKLT model for which there is a unique ground state. Even in this case, we find that the
negativity of separated blocks of spins is zero. 
\end{abstract}

\maketitle

During the last years, in the disciplines of quantum information theory, quantum many-body physics and statistical mechanics, the study of 
purely quantum effects like entanglement has become important. The idea of entanglement was first introduced by Schr\"{o}dinger \cite{Sch} in
the early days of quantum mechanics. 
Entanglement is a phenomena where two (or more) quantum systems are linked together and their description cannot be done separately, disregarding 
their spatial separation.

Entanglement plays a fundamental role in some quantum mechanical systems, being important in the description of quantum phase 
transitions, \cite{Amico}, topological order \cite{Kitaev1,Levine} and even macroscopic properties of solids \cite{Ghosh}.
For pure systems, the measure of bipartite entanglement is given by the entanglement entropy, or von Neumann entropy, defined as 
$S[\rho]={\rm Tr\rho\ln\rho}$, where the state is characterized through it's density matrix $\rho$. 

Unfortunately for mixed states, the von Neumann entropy is not
an appropriate measure of quantum entanglement. One is thereby forced
to use other measures
of entanglement for mixed states \cite{Eisert,Horodecki}. In this letter we focus on a
particular measure of entanglement called the negativity, which stems
from the Peres separability condition \cite{P}, and first proven to be a
bona fide quantifier of entanglement in \cite{8a}. If one is to quantify the
genuine quantum correlations or entanglement between {\em
non-complementary} parts of a many body system (as opposed to mutual
information, which
only quantifies {\em total} correlations), one has to resort to a measure
such as negativity. Unfortunately, such calculations have been found to be only
numerically tractable so far \cite{8b}, except for
infinite range models \cite{8c}.

Given a system $E$, we define two subsystems $A$ and $B$, characterized by the density matrix ${\rm Tr}_{C}\rho=\rho_{AB}$, where we have
traced away the degrees of freedom $C$, who lie outside the $A$ and $B$ subsystems. Negativity is defined then as the sum of negatives 
eigenvalues of the partial transposed density matrix $\rho^{T_A}_{AB}$, where the transposition is done in the $A$ subsystem.
The existence of nonzero negativity is signal of entanglement, but, as pointed out in \cite{Horodecki2}, vanishing negativity does not imply
zero entanglement. 

In this paper, we present the result for the negativity of two blocks defined as subsystems of the AKLT ground state.
In quantum information, entanglement between separated systems is a
valuable resource. In this sense it is important to calculate the
entanglement between separated (i.e. non-complementary) blocks of a
system. It is therefore important to examine whether such an
entanglement resource is present in the AKLT model whose ground state
has already been demonstrated to be an ideal resource for measurement
based quantum wires \cite{miyake} in 1D, and quantum
computation \cite{Wei} in 2D. It is a realistic model of spin systems
in the sense of being
 short range and models the gapped nature of integer spin models.
Perhaps its most attractive feature is its analytic solvability and
thereby it holds the promise of the first ever analytic computation of
negativity for a short range model.

\section*{AKLT model}

We will study two different scenarios for the problem. In the first case, we will compute the negativity in the AKLT model for
a bulk of spins 1 and one spin 1/2 at each boundary. This system has a unique ground state, as was shown in \cite{KK} for the general case.
In the second scenario, we study the case of a bulk \textit{and} the boundary made up of spins 1. This system should be easier to realize 
experimentally than the previous one, but it has some subtleties as it's ground state is four fold degenerate.

\subsection{Spin 1/2 at the boundary}

The one dimensional AKLT model \cite{AKLT0} that we will consider consists of a chain of $N$ spin-$
1$’s in the bulk, and two spin-$1/2$ on the boundary. The location where the spins sit are called sites. We
shall denote by $\vec{S}_k$ the vector of spin-$1$ operators and by
$\vec{s}_b$ spin$-1/2$ operators, where $b = 0, N + 1$. The Hamiltonian is:
       
\begin{equation}\label{AKLT_ham}
 H=H_{\rm Bulk}+\Pi_{0,1} + \Pi_{N,N+1},
\end{equation}

\noindent where the Hamiltonian corresponding to the bulk is given by

\begin{eqnarray}\label{AKLT_bulk}
 H_{\rm Bulk}&=&\sum_{i=1}^{N-1}P(\vec{S}_i+\vec{S}_{i+1}).
\end{eqnarray}

\noindent Here $P(\vec{S}_i+\vec{S}_{i+1})$ is a projector onto spin 2 states, given by
\begin{equation}
P(\vec{S}_k+\vec{S}_{k+1})=\frac{1}{6}\left(3\vec{S}_k\cdot\vec{S}_{k+1} + (\vec{S}_k\cdot \vec{S}_{k+1})^2 +2\right),
\end{equation}

\noindent and the sum runs over the lattice sites. The boundary terms $\Pi$ describe interaction of a spin $1/2$
and spin $1$. Each term is a projector on a state with spin $3/2$:

\begin{equation}
 \Pi_{0,1}=\frac{2}{3}(1+\vec{s}_0\cdot\vec{S}_1), \quad \Pi_{N,N +1} =\frac{2}{3}(1 + \vec{S}_N \cdot\vec{s}_{N+1}).
\end{equation}

In order to construct the ground state $|{\rm VBS}\rangle$ of (\ref{AKLT_ham}) we can associate two spin $1/2$ variables at each lattice site 
and create the spin $1$ state symmetrizing them. To prevent the formation of spin $2$, we antisymmetrize states between 
different neighbor lattice sites.  Doing this we are sure that this configuration is actually an eigenstate of 
the Hamiltonian (\ref{AKLT_ham}), with eigenvalue $0$ (i.e. the projection of $|{\rm VBS}\rangle$ on the subspace of spin 2-states is zero). Noting that
the Hamiltonian (\ref{AKLT_ham}) is positive definite, then we know that this is the ground state. It is possible to write down a compact expression
for this VBS state using bosonic variables. Following \cite{AAH}, we make use of 
the Schwinger boson representation for $SU(2)$ algebra at each site $j$, namely:

\begin{eqnarray}
S^+_j=a^\dagger_j b_j,\quad S^-_{j}=a_jb^\dagger_j,\quad S^z_{j}=\frac{1}{2}(a_j^\dagger a_j-b_j^\dagger b_j),\\\nonumber
\mbox{with}\quad [S^z_i,S^\pm_j]=\pm S^\pm_i \delta_{ij}, \quad [S^+_i,S^{-}_j]=+2 S^z_i \delta_{ij},
\end{eqnarray}

\noindent where $a$ and $b$ are two sets of bosonic creation operators, with the usual commutation relations 
$[a_i,a^\dagger_j]=[b_i,b^\dagger_j]=\delta_{ij}$, $[a_i,a_j]=[b_i,b_j]=0$ and correspondingly for $a^\dagger$ and $b^\dagger$. 
This two sets commute in each and every lattice site, i.e. $[a_i,b_j]=[a_i^\dagger,b_j]=0$. In terms of these variables, the ground state can be 
written as

\begin{equation}\label{VBS_state}
 |{\rm VBS}\rangle=\prod_{i=0}^N(a_i^\dagger b_{i+1}^\dagger - a_{i+1}^\dagger b_i^\dagger)|0\rangle.
\end{equation}

\noindent where $|0\rangle=\bigotimes_{sites}|0_a,j\rangle\otimes|0_b,j\rangle$. The state $|0_a,j\rangle$ is defined by $a_j|0_a,j\rangle=0$, 
and it's called the vacuum state for the set of operators $a$. $|0_b,j\rangle$ is defined similarly for the set $b$.
In \cite{KK} the authors prove that this ground state is unique for the Hamiltonian (\ref{AKLT_ham}), then we can construct
the density matrix of the (pure) ground state

\begin{equation}\label{density_M}
 \rho=\frac{|{\rm VBS}\rangle\langle {\rm VBS}|}{\langle {\rm VBS}|{\rm VBS}\rangle}.
\end{equation}

This is a one dimensional projector on the ground state of the Hamiltonian (\ref{AKLT_ham}).

The bulk Hamiltonian (\ref{AKLT_bulk}) possess a ground state which is four-fold degenerate. We label the different orthogonal ground states
by a Greek letter, which can take the values $\mu=0..3$. This ground states are defined by the action of four operators $T_\mu$, which act on the 
boundary of a state defined in the bulk in a similar way as (\ref{VBS_state}). This state has spin 1 at each lattice site, but spin 1/2 at the 
boundary, then the action of the $T_\mu$ operators is to create spin 1 also at the boundary.
We have, for the Hamiltonian (\ref{AKLT_bulk})

\begin{eqnarray}\nonumber
|{\rm GS}_\mu\rangle&=& T_\mu^\dagger(1,N-1)|{\rm GS}\rangle\\
&=&T_\mu^\dagger(1,N-1)\prod_{i=1}^{N-1}(a_i^\dagger b_{i+1}^\dagger - a_{i+1}^\dagger b_i^\dagger)|0\rangle.
\end{eqnarray}

The $T_\mu$ operators can be defined in term of the boson creation operators $a$ and $b$ as

\begin{eqnarray}\nonumber\label{T_op}
 T_0^\dagger(i,j)=a^\dagger_ia^\dagger_j +b^\dagger_ib^\dagger_j, \quad T_1^\dagger(i,j)=a^\dagger_ib^\dagger_j +b^\dagger_ia^\dagger_j,\\
 T_2^\dagger(i,j)=i(a^\dagger_ib^\dagger_j -b^\dagger_ia^\dagger_j), \quad T_3^\dagger(i,j)=a^\dagger_ia^\dagger_j -b^\dagger_ib^\dagger_j.
\end{eqnarray}

\noindent then, we have $ H_{\rm Bulk}|{\rm GS}_\mu\rangle=0,$ $(\mu=0,1,2,3).$ Linear combinations of this operators
acting in the boundary of the chain can create separable states (take for example $T^\dagger_0+T^\dagger_3$). Such states are also valid ground states of
the chain. We will elaborate more on this point in section \ref{Spin1}.

\section*{Negativity for the mixed system of 2 blocks}\label{N2systems}

\begin{center}
\begin{figure}[ht!]
 \includegraphics[scale=.58]{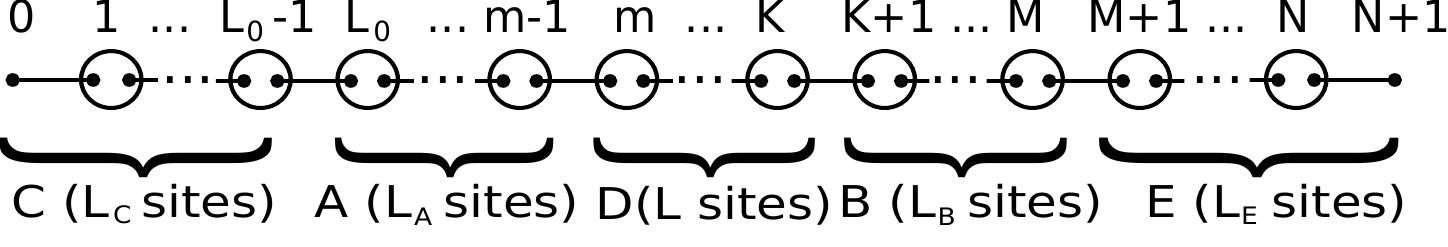}
 \caption{We made a partition of the VBS state in 5 sectors, labeled $A,B,C,D$ and $E$ as shown in the figure. To obtain the density 
matrix for the blocks $A$ and $B$, we trace away the spin variables at the sites inside $C,D$ and $E$.}\label{fig:mutualp}
\end{figure}
\end{center}

We study the mixed system composed of two blocks $A$ and $B$ of length $L_A$ and $L_B$, obtained by tracing away the lattice sites which 
do not belong to these blocks in the VBS ground state. This situation is described in Fig \ref{fig:mutualp}.

\subsection{Spin 1/2 at the boundary}\label{1/2boundary}

To define the blocks, we partition the $N+2$ sites of the chain into five different subsets, $A,B,C,D$ and $E$, of different length.

Given $I,J,K,M,N$ five positive integers ordered as $0< I < J < K < M < N+1$, we define:

\begin{itemize}
 \item Block C = $\{$sites $i$, $0 \leq i \leq I-1\}$, with length $L_C=I$,
 \item Block A = $\{$sites $i$, $I \leq i \leq J-1\}$, with length $L_A=J-I$,
 \item Block D = $\{$sites $i$, $J \leq i \leq K-1\}$, with length $L=K-J$,
 \item Block B = $\{$sites $i$, $K \leq i \leq M-1\}$  with length $L_B=M-K$ and 
 \item Block E = $\{$sites $i$, $M \leq i \leq N+1\}$  with length $L_E=N+2-M$.
\end{itemize}

We are interested in the density matrix for the mixed system of $A$ and $B$ blocks. We obtain this density matrix by tracing away the states 
on the $C,D$ and $E$ subspaces.

\begin{equation}\label{matrix_mix}
 \rho_{AB}={\rm Tr}_{CDE}(\rho)
\end{equation}

We can write the VBS state as a linear combination of products between the different four 
fold degenerate ground states of the bulk Hamiltonian (\ref{AKLT_bulk}) in the form: (implicit summation assumed)

\begin{flalign}\nonumber
&|{\rm VBS}\rangle=M_{\mu\nu\rho\sigma} T_\sigma^\dagger(I-1,M)|C,A_\mu,D_\nu,B_\rho,E\rangle,\quad \mbox{with}&\\
&M_{\mu\nu\rho\sigma}=(-1)^\nu(\delta_{\mu}^{\nu}g_{\rho\sigma}+\delta^{\nu}_{\rho}g_{\mu\sigma}-\delta_{\sigma}^{\nu}g_{\mu\rho}+
ig^{\nu\alpha}\epsilon_{\mu\alpha\rho\sigma}).&
\end{flalign}

\noindent here we have introduced three types of tensors, the Kronecker delta symbol in 4 dimensions $\delta_\alpha^\beta$, the diagonal tensor 
$g_{\mu\nu}=g^{\mu\nu}={\rm diag}(-1,+1,+1,+1)$ and the Levi Civita tensor in four dimensions $\epsilon_{\mu\nu\rho\sigma}$, which is a totally antisymmetric tensor, 
with $\epsilon_{\mu\nu\rho\sigma}$= sign of permutation $(\mu,\nu,\rho,\sigma)$ if $(\mu,\nu,\rho,\sigma)$ is a permutation of $(0,1,2,3)$, 
and zero otherwise.

Using this representation of the VBS state, it is easy to write down the density matrix (\ref{matrix_mix}) using the 
{or}\-{tho}\-{go}\-{na}\-{li}\-{ty} of the bulk ground states, namely

\begin{flalign}
 &\langle D_\mu|D_\nu\rangle=\delta^{\mu}_\nu\lambda_\mu(L)\quad \mbox{with } \lambda_\mu(L)=\frac{1}{4}+\frac{z(L)}{4}s_\mu,&\\\nonumber
&z(L)=\left(-\frac{1}{3}\right)^L;  \quad s_\mu=(-1,-1,3,-1), \\
&\langle C,E]T_\mu(I-1,M){T_\nu}^\dagger(I-1,M)\left[C,E\right\rangle=\delta^{\mu}_\nu.&
\end{flalign}

\noindent we find that the density matrix $\rho_{AB}$ is

\begin{equation}\label{dens_matrix_mix}
\rho_{AB}= M_{\mu\nu\rho\sigma} M_{\alpha\nu\beta\sigma}|A_\mu,B_\rho,\rangle\langle A_\alpha,B_\beta|,
\end{equation}

\noindent with the tensor $M_{\mu\nu\rho\sigma}M_{\alpha\nu\beta\sigma}$ given explicitly by  (summation over dummy variables $\nu$ 
and $\sigma$ is assumed)

\begin{flalign}\nonumber\label{dens_matrix_mix2}
 M_{\mu\nu\rho\sigma}M_{\alpha\nu\beta\sigma}=\delta_{\mu}^{\alpha}\delta_{\rho}^{\beta}+z(L)[\delta_{\mu}^{\rho}\delta_{\alpha}^{\beta}-g^{\rho\alpha}g^{\mu\beta}]S_{\mu\alpha}\\
+iz(L)g^{\lambda\alpha}g^{\sigma\beta}\epsilon_{\mu\rho\lambda\sigma}\left(S_{\rho\beta}-S_{\mu\alpha}\right),
\end{flalign}

\noindent with $S_{\mu\alpha}=(s_\mu+s_\alpha)/{2}$. This is the first explicit form of the reduced density matrix of two non complementary blocks
for the AKLT model. This is important from the point of view of the fact that recently obtaining the reduced density matrices of non-complementary blocks in a many-body system
has been the focus of much interest \cite{calabrese}.
We can identify two parts in (\ref{dens_matrix_mix}), the first term which does not depend on 
$z$ and the rest which is linear in $z$. The first term is a projector on the ground states of the bulk of $A$ and $B$, namely

\begin{equation}\label{rho0}
 \rho_0(A,B)=\delta_{\mu}^{\alpha}\delta_{\rho}^{\beta}|A_\mu,B_\rho,\rangle\langle A_\alpha,B_\beta|,
\end{equation}

\noindent while all the other terms, proportional to $z(L)$, have vanishing trace. If we call $\rho_1(A,B)$ to all the linear terms in $z(L)$ on 
(\ref{dens_matrix_mix}), we can write for brevity

\begin{equation}\label{breve}
\rho_{AB}= \rho_0(A,B)+z(L)\rho_1(A,B).
\end{equation}

From the expressions (\ref{dens_matrix_mix}) and (\ref{dens_matrix_mix2}) we can obtain the partial transposed density matrix with respect to 
the $A$ subsystem.

\begin{flalign}\label{partialtrans1}
&\rho^{T_{A}}_{AB}=\bigg[\delta_{\mu}^{\alpha}\delta_{\rho}^{\beta}
+z(L)[\delta_{\alpha}^{\rho}\delta_{\mu}^{\beta}-g^{\rho\mu}g^{\alpha\beta}]S_{\mu\alpha}&\\\nonumber
&+iz(L)g^{\lambda\mu}g^{\sigma\beta}\epsilon_{\alpha\rho\lambda\sigma}(S_{\rho\beta}-S_{\mu\alpha})\bigg]|A_\mu,B_\rho,\rangle\langle A_\alpha,B_\beta|.&
\end{flalign}

If we perform a unitary transformation $U$ on the basis vectors $|A_\mu,B_\rho\rangle$ defined by it's action on the basis as
$U|A_\mu,B_\rho\rangle=g^{\mu\sigma}|A_\sigma,B_\rho\rangle$, we find

\begin{flalign}
&U\rho^{T_{A}}_{AB}U^\dagger=\bigg[\delta_{\mu}^{\alpha}\delta_{\rho}^{\beta}
-z(L)[\delta_{\mu}^{\rho}\delta_{\alpha}^{\beta}-g^{\rho\alpha}g^{\mu\beta}]S_{\mu\alpha}&\\\nonumber
&-iz(L)g^{\lambda\alpha}g^{\sigma\beta}\epsilon_{\mu\rho\lambda\sigma}(S_{\rho\beta}-S_{\mu\alpha})\bigg]|A_\mu,B_\rho,\rangle\langle A_\alpha,B_\beta|,&
\end{flalign}

\noindent from where, comparing with equations (\ref{dens_matrix_mix}) and (\ref{dens_matrix_mix2}), we learn that 

\begin{equation}\label{equivalence}
U\rho^{T_{A}}_{AB}(z)U^\dagger=\rho_{AB}(-z)
\end{equation}

With this result, we can state our main theorem:

\begin{Theorem1}
The negativity of the transposed density matrix $\rho^{T_{A}}_{AB}(z(L))$ is strictly zero for two blocks separated by $L>0$.
\end{Theorem1}
\begin{proof} 
Consider the family of density matrices $\rho_{AB}(z)=\rho_0(A,B)+z(L)\rho_1(A,B)$,  defined in eq. (\ref{breve}). Recalling
that the space of density matrices is convex \cite{Nielsen}, meaning that for two density matrices $\rho_1,\rho_2$, the operator
$\tilde{\rho}=\lambda\rho_1+(1-\lambda)\rho_2$ is also a density matrix for $\lambda\in[0,1]$, we proceed as follows.
We take the first two members of the family $\rho_{AB}(z)$, namely $\rho_{AB}(z_1)$ and $\rho_{AB}(z_2)$ for fixed $z_1=1,z_2=-1/3$ \footnote{ 
We can take any pair different $z_1$ and $z_2$, but the greater $z$ is achieved for $z_1=1$, $z_2=-1/3$ (or vice versa)}. 
By the convexity of the space of density matrices, $\bar{\rho}=\lambda\rho_{AB}(z_1)+(1-\lambda)\rho_{AB}(z_2)$ is also a density
matrix. Using (\ref{breve}), we write explicitly $\bar{\rho}=\rho_0(A,B)+(\lambda z_1 + (1-\lambda)z_2)\rho_1(A,B)$. We can choose
$\lambda=\frac{1}{4}(1-3(-\frac{1}{3})^L)\in [0,1]$ for $L\geq 1$. Using this $\lambda$, we find 

\begin{equation}
 \bar{\rho}=\rho_0(A,B)-z(L)\rho_1(A,B).
\end{equation}

Then $\bar{\rho}=\rho_{AB}(-z)$ is also a density matrix, for $L\geq 1$. Now, by (\ref{equivalence}), $\rho^{T_{A}}_{AB}(z)$
is also density matrix for $z<1$ $(L>0)$. Then the negativity (sum of negative eigenvalues) of $\rho^{T_{A}}_{AB}(z)$ vanish.
\end{proof}

For the case when the blocks are adjacent ($L=0$), the negativity in the limit $L_A,L_B\gg1$ is

\begin{equation}
 N(A,B)=\frac{1}{2} -\frac{3}{4}(z(L_A)^2+z(L_B)^2).
\end{equation}
This means that as blocks get large, if they are touching, then their entanglement approaches that of a maximally entangled pair of q-bits. 

It is pertinent here to provide an alternative proof of the vanishing negativity of non-complementary blocks in the above AKLT ground state (i.e. with spin-1/2s at the ends) which may be obtained 
by exploiting the property that negativity is an entanglement monotone \cite{Vidal}, as well as the projective (filtering) mechanism through which the ground state of a larger AKLT chain can be grown from the ground state of a smaller chain. For this, first consider the 3 spin version of the above AKLT chain which has the ground state
\begin{eqnarray}
(a_0^{\dagger}b_1^{\dagger}-a_1^{\dagger}b_0^{\dagger})(a_1^{\dagger}b_2^{\dagger}-a_2^{\dagger}b_1^{\dagger})|0\rangle \nonumber \\=-\sqrt{\frac{1}{3}}|+\frac{1}{2}\rangle_0 |-1\rangle_1|+\frac{1}{2}\rangle_2+\nonumber\\ \sqrt{\frac{1}{6}}|-\frac{1}{2}\rangle_0 |0\rangle_1|+\frac{1}{2}\rangle_2+\sqrt{\frac{1}{6}}|\frac{1}{2}\rangle_0 |0\rangle_1|-\frac{1}{2}\rangle_2
\nonumber\\- \sqrt{\frac{1}{3}}|-\frac{1}{2}\rangle_0 |+1\rangle_1|-\frac{1}{2}\rangle_2.
\end{eqnarray}
The $4\times 4$ reduced density operator of the spins at sites $0$ and $2$ (both being spin-1/2 particles) is then
\begin{eqnarray} 
\rho_{02}=\frac{1}{3}(|+\frac{1}{2}, +\frac{1}{2}\rangle\langle +\frac{1}{2}, +\frac{1}{2}|+|-\frac{1}{2}, -\frac{1}{2}\rangle\langle -\frac{1}{2}, -\frac{1}{2}|\nonumber\\+ |\psi^{+}\rangle\langle\psi^{+}|),
\end{eqnarray}
where $|\psi^{+}\rangle=\frac{1}{\sqrt{2}}(|+\frac{1}{2}\rangle|-\frac{1}{2}\rangle+|-\frac{1}{2}\rangle|+\frac{1}{2}\rangle)$. 
It is easy to check that $\rho_{02}$ has a vanishing negativity.
Now, the ground state of the 4 spin version of the AKLT chain with spin-1/2s at the ends may be obtained from the 3 spin version by 
bringing in a pair of spin-1/2 particles $2'$ and $3$ in a singlet state 
$|\psi^{-}\rangle=\frac{1}{\sqrt{2}}(|+\frac{1}{2}\rangle|-\frac{1}{2}\rangle-|-\frac{1}{2}\rangle|+\frac{1}{2}\rangle)$ and projecting 
$2$ and $2'$ to their symmetric subspace to make a new 
spin-1 particle at site $2$. After this operation, sites $0$ and $3$ have spin-1/2 particles, while sites $1$ and $2$ have spin-1 
particles and the system as a whole is the ground state of the 4 spin AKLT model with spin-1/2s at the boundaries.  The block composed of 
spins 2 and 3 has thereby arisen from the spin 2 by a local action (a positive operator valued measurement) and a selection of a certain 
outcome (in this case a projection to a certain subspace). Since the negativity between $0$ and $2$ in $\rho_{02}$ extracted from the 3 
spin AKLT model is zero, then the negativity between $0$ and the block composed of spins $2$ and $3$ in the 4 spin AKLT model should also 
vanish by virtue of the fact that negativity is an entanglement monotone (Eq.(17) of Ref.\cite{Vidal}).
Continuing by induction, adding spins on both sides of the chain to construct AKLT ground states with larger numbers of spins, it may thereby be shown that for blocks not touching each other the negativity vanishes.

\subsection{Spin 1 at the boundary}\label{Spin1}

\begin{center}
\begin{figure}[ht!]
 \includegraphics[scale=.45]{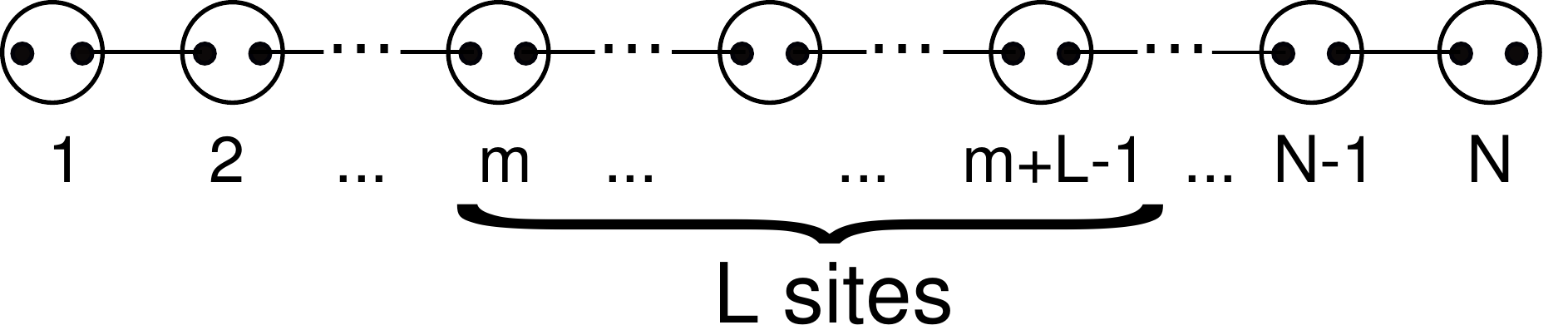}
 \caption{Starting from one of the four ground states of the VBS chain, we trace away one block of $L$ sites. This creates a mixed state
of two blocks.}\label{fig:VBS4}
\end{figure}
\end{center}

The Hamiltonian in this case is just (\ref{AKLT_bulk}) with no boundary terms. The ground state of a chain of spins 1, is fourfold degenerate 
as shown in \cite{AKLT0}. We label these ground states by a greek index, which can take values from zero to three. We have

\begin{equation}
 |{\rm VBS}_\mu\rangle=T_\mu^\dagger(1,N)\prod_{i=1}^N(a^\dagger_ib_{i+1}^\dagger-a^\dagger_{i+1}b_{i}^\dagger)|0\rangle.
\end{equation}

\noindent The $T_\mu$ operators are defined as in (\ref{T_op}). We define the states in this problem as $A=1 \leq{\rm sites}\leq m-1$,
 $B=m \leq{\rm sites}\leq m+L-1$ and  $C=m+L \leq{\rm sites}\leq N$,

\begin{eqnarray}
|O\rangle=\prod_{i,i+1\in O}(a^\dagger_ib_{i+1}^\dagger-a^\dagger_{i+1}b_{i}^\dagger)|0\rangle.
\end{eqnarray}

The ${\rm VBS}$ state splits up in 

\begin{eqnarray}\nonumber
 |{\rm VBS}_\rho\rangle&=&T_\rho^\dagger(1,N)T_2^\dagger(m-1,m)T_2^\dagger(K,K+1)|A,B,C\rangle,\\
&=& \frac{1}{2}g^{\mu\nu}T_\rho^\dagger(1,N)T_\mu^\dagger(m-1,K+1)|A,B,C_\nu\rangle\label{VBS_mu}
\end{eqnarray}

\noindent where $K=m+L-1$, and $|A,B,C\rangle=|A\rangle|B\rangle|C\rangle$.

The density matrix of the $A,B$ subsystems can be computed directly from the expression (\ref{VBS_mu}), using the orthogonality of the bulk 
states $|C_\nu\rangle$. Defining $|A,B\rangle_\mu\equiv T_\mu^\dagger(m-1,K+1)|A,B\rangle$, we have

\begin{eqnarray}
\rho_{AB}(\mu)=\lambda_\mu T_\rho^\dagger(1,N)|A,B\rangle_\mu\langle A,B|_\mu T_\rho(1,N).
\end{eqnarray}

We can write this density matrix in terms of the ground states of the $A$ and $B$ bulks. In terms of these we have (no implicit sum over $\beta$)

\begin{eqnarray}\nonumber
\rho_{AB}(\beta)=\lambda_\alpha[\sigma_\lambda\sigma_\alpha\sigma_\mu\sigma_\beta]\overline{[\sigma_{\lambda'}\sigma_\alpha\sigma_{\mu'}\sigma_\beta]}|A_\mu,B_\lambda\rangle\langle A_{\mu'},B_{\lambda'}|,
\end{eqnarray}

\noindent $(\lambda_\alpha=\lambda_\alpha(L)$), with $[\sigma_\lambda\sigma_\alpha\sigma_\mu\sigma_\beta]={\rm Trace}(\sigma_\lambda\sigma_\alpha\sigma_\mu\sigma_\beta)$ and $\overline{a}$
being the complex conjugate of $a$. From this expression we can easily find the transposed density matrix respect to the $A$ subsystem. We have

\begin{flalign}\label{trans_mat}
\rho_{AB}^{T_A}(\beta)=\lambda_\alpha[\sigma_\lambda\sigma_\alpha\sigma_\mu\sigma_\beta]\overline{[\sigma_{\lambda'}\sigma_\alpha\sigma_{\mu'}\sigma_\beta]}|A_{\mu'},B_\lambda\rangle\langle A_{\mu},B_{\lambda'}|.
\end{flalign}

The negativity of the system is given by the sum of negative eigenvalues of (\ref{trans_mat}). Taking $L_A,L_B$ and $L$ the length of the blocks $A$, $B$
and $C$ respectively, we have in the limit of big blocks

\begin{eqnarray}
&N_{L=0}=\frac{3}{2},\quad N_{L>0}=\frac{1}{2} \quad\mbox{for} \quad L_A,L_B=\infty,&\\\nonumber
&N=\frac{1}{2}-\frac{3}{4}(z(L_A)^2+z(L_B)^2) \,\,\, \mbox{for} \,\,\, L_A,L_B,L\gg 1.&
\end{eqnarray}

\noindent  Another interesting limit is when one block is infinitely long compared with the other ($L_A=\infty,L_B=1$ for any separation $L$)

\begin{eqnarray}\nonumber
N&=&\frac{1}{24}(3\sqrt{9-10z(L)+17z(L)^2}+5z(L)-1),\\
&\simeq& \frac{1}{3} +\frac{8}{27}z(L)^2 \,\,\, \mbox{for} \,\,\,L\gg1.
\end{eqnarray}

In the thermodynamic limit $L_A,L_B=\infty$ (where the four ground states become indistinguishable\cite{AKLT0}), the negativity quite surprisingly 
does not depend on the separation of the blocks. The topology of the system determines the entanglement properties. Then when the blocks are 
adjacent (i.e $L=0$), the negativity is maximal being $N=3/2$, while for any other separation
$N=1/2$.

To obtain the previous result, we started from (\ref{VBS_mu}), which is a entangled state between the first and the last spin. Taking linear 
combinations of $|{\rm VBS}\rangle$
states, we can construct unentangled states between the first and the last spins. Relabelling $\psi^1_i=a_i,\psi^2_i=b_i$, we can write a general 
unentangled state between the first and the last spin 1/2 in the chain as

\begin{equation}
 |{\rm ^{cd}VBS}\rangle={\psi_1^\dagger}^c{\psi_N^\dagger}^d\prod_{i=1}^N(a^\dagger_ib_{i+1}^\dagger-a^\dagger_{i+1}b_{i}^\dagger)|0\rangle.
\end{equation}

\noindent This states can be obtained as linear combination of the states $|{\rm VBS}_\mu\rangle$, for example, the state 
$|^{11}{\rm VBS}\rangle=|{\rm VBS}_0\rangle+|{\rm VBS}_3\rangle$, and so on.
The density matrix in this case, after tracing out block $C$ is $(c,d=1,2)$

\begin{eqnarray}\label{rho_un}\nonumber
\rho_{AB}(c,d)&=&\lambda_\mu(L){\psi_1^\dagger}^c{\psi_N^\dagger}^d|A,B\rangle_\mu\langle A,B|_\mu{\psi_1}^c{\psi_N}^d\\
&\equiv&\lambda_\mu(L)|^{cd}A,B\rangle_\mu\langle ^{cd} A,B|_\mu.
\end{eqnarray}

\noindent The partial transposed density matrix can be calculated directly from (\ref{rho_un}), the result is

\begin{equation}\label{rho_un}
 \rho_{AB}(c,d)=u_\mu(L)|^{cd}A,B\rangle_\mu\langle ^{cd} A,B|_\mu.
\end{equation}

\noindent with $u_\mu(L)=(\lambda_\mu(L)-\frac{1}{2}(-1)^\mu(\lambda_2(L)-\lambda_1(L))$. Using this result, we compute the negativity of
the system which turns out to vanish for $L\neq0$, while for $L=0$, we have

\begin{flalign}\nonumber
N(A,B;c,d)=\frac{\sqrt{1+z(L_A)^2z(L_B)^2-z(L_A)^2-z(L_B)^2}}{2-2\phi^{cd}z(L_A)z(L_B)},\\
\approx\frac{1}{2}-\frac{1}{4}(z(L_A)+\phi^{cd}z(L_B))^2 \,\,\, (L_A,L_B\gg 1).
\end{flalign}

\noindent where we have introduced the symbol $\phi^{cd}$ which is equal to $1$ if $c=d$ and takes the value
$-1$ if $c\neq d$.

\subsection{Periodic boundary conditions}\label{PBC}

Another case widely studied is the scenario with periodic boundary conditions. We can obtain this state from the case studied in the previous 
section, antisymmetrizing the first and the last free spin 1/2 variables. This state is unique, as follows given that the coordination number for 
each spin is two \cite{KK}. 

\begin{center}
\begin{figure}[ht!]
 \includegraphics[scale=.7]{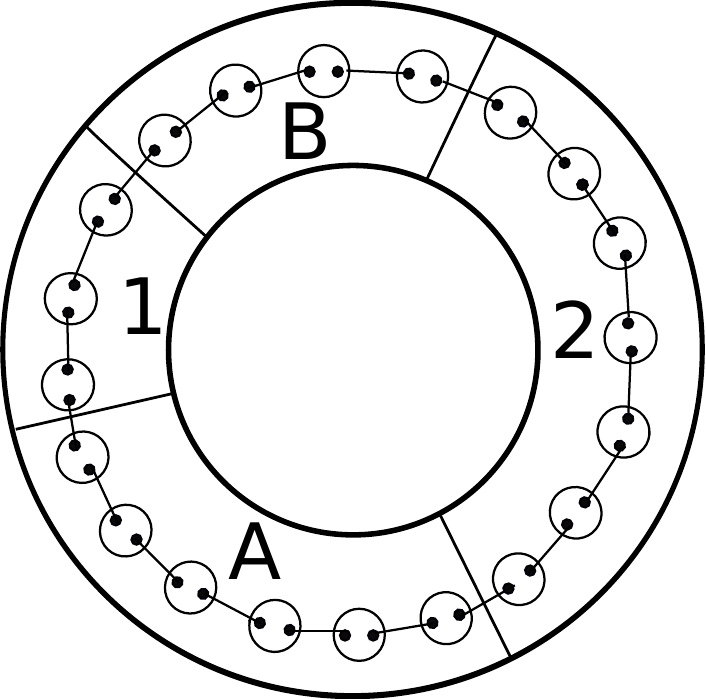}
 \caption{We trace blocks 1 and 2, leaving a reduced density matrix in terms of the states of the A and B blocks.}\label{fig:bc}
\end{figure}
\end{center}

In this state, we make a partition in four sectors, labeled by their length as $L_1, L_A, L_2, L_B$, with $L_1+L_2+L_A+L_B=L$ the total length 
of the system. We trace away the states from the sectors that do not belong to $A\cup B$ (See fig. \ref{fig:bc}). 

The reduced density matrix in this case is given by

\begin{eqnarray}
\rho_{AB}=M_{\alpha\beta\alpha'\beta'}|A_\alpha,B_\beta\rangle\langle A_{\alpha'},B_{\beta'}|.
\end{eqnarray}

The tensor $M_{\alpha\beta\alpha'\beta'}$ is given explicitly by

\begin{eqnarray}\nonumber
M_{\alpha\beta\alpha'\beta'}=\delta_{\alpha}^{\alpha'}\delta_{\beta}^{\beta'}\Lambda_{\alpha\beta}(z_1,z_2)-g_{\alpha\beta}g^{\alpha'\beta'}\Gamma_{\alpha\alpha'}(-z_1,-z_2)\\\nonumber
+ig^{\lambda\beta'}g^{\mu\alpha'}\epsilon_{\beta\lambda\alpha\mu}(T_{\alpha\beta\alpha'\beta'}(z_1,z_2)-\delta_{\alpha}^{\beta'}\delta^{\beta}_{\alpha'}\Gamma_{\alpha\alpha'}(z_1,z_2),\\
\end{eqnarray}

\noindent where $z_1=z(L_1)$ and $z_2=z(L_2)$. The tensors $\Lambda_{\alpha\beta}(x,y),\Gamma_{\alpha\alpha'}(x,y)$ and $T_{\alpha\beta\alpha'\beta'}(x,y)$
are respectively given by 

\begin{eqnarray}\nonumber
\Lambda_{\alpha\beta}(x,y)=\frac{1+(s_\alpha s_\beta+s_\alpha+s_\beta)xy}{1+3z(L)},\\\nonumber
\Gamma_{\alpha\alpha'}(x,y)=\frac{s_\alpha+s_{\alpha'}}{1+3z(L)}\left(xy-\frac{x+y}{2}\right) \quad\mbox{and} \\
T_{\alpha\beta\alpha'\beta'}(x,y)=\frac{s_\alpha-s_\beta+s_{\alpha'}-s_{\beta'}}{4+12z(L)}(x-y).
\end{eqnarray}

As with the case studied in section \ref{1/2boundary}, it is easy to see that if we write the partial density matrix in the form 
$\rho_{AB}(z_1,z_2)=\rho_0+z_1\rho_1+z_2\rho_2+z_1z_2\rho_3$ the partial transposed density matrix $\rho_{AB}^{T_A}$ is isomorphic (up to change of
basis) to $\rho_{AB}(-z_1,-z_2)$. We can use again the convexity of the space of density matrices to prove that the matrix $\rho_{AB}(-z_1,-z_2)$
is also a density matrix, for $L_1,L_2\geq 1.$ From this result we see that negativity vanish for $L_1,L_2\geq 1$.

\section*{Conclusions}

Understanding the entanglement structure of AKLT states is important because AKLT states have recently been recognized as useful 
resources for quantum information processing \cite{miyake,Wei}. We have computed the entanglement between non-complementary 
blocks of spins in such a system. We have found that the AKLT chain allows an analytic computation of the negativity
between non-complementary blocks in a short range (realistic) spin model, which has not been accomplished in the other models
studied to date.

The entanglement properties of a system of two blocks made up from the ground state of an AKLT system depends dramatically on whether
the blocks touch each other. 
We have proved that the negativity of a system of two blocks separated by $L\geq1$ sites obtained from a unique ground state (\ref{VBS_state}), 
is strictly zero. 

In the case when all the sites contain a spin 1 and open boundary conditions, the entanglement properties, and particularly the negativity, depends on the configuration between
the sites at the boundary. Each of the spin 1 variables can be constructed from the symmetrization of two spin 1/2 states. In particular, the sites at the boundary can be 
constructed through this process. In this boundary sites, however, one of those spin 1/2 can be in any state, i.e. $|\uparrow\rangle$ or 
$|\downarrow\rangle$. We can construct Bell states between the first and the last spins 1/2 in the chain. This states have non vanishing negativity
regardless the separation between the subsystems $A$ and $B$ inside the chain. If, on the other hand, the spins 1/2 at the end of the chain are 
in one of the four separable states (i.e. calling the first and the last site $1$ and $N$ respectively, $|\uparrow_1\uparrow_N\rangle$,
$|\uparrow_1\downarrow_N\rangle$,$|\downarrow_1\uparrow_N\rangle$,$|\downarrow_1\downarrow_N\rangle$) then the negativity vanish when the subsystems
$A$ and $B$ are separated. Thus, in practice, as there is no easy way to isolate one of those AKLT ground states which have a Bell state between its end spins, we would
expect vanishing negativity even for the case when all sites contain a spin 1.  

 We have also shown that the vanishing negativity between separated blocks of spins also holds for the periodic AKLT  which has a unique ground state. Other important results of
our work include a general expression for the reduced density matrix of an arbitrary pair of blocks of spins in a AKLT chain that includes the case of separated blocks, and the negativities
between those non-complementary blocks that touch each other at a point.

\smallskip

\textbf{Acknowledgments} R. S. acknowledges the Fulbright-Conicyt Fellowship. V.K. acknowledges the Grant DMS-0905744.

\end{document}